\documentclass[conference,9pt]{IEEEtran}
\IEEEoverridecommandlockouts 

\usepackage[utf8]{inputenc} 
\usepackage{pifont} 
\usepackage{xcolor}
\usepackage{times}
\usepackage{graphicx}
\usepackage{algorithm}
\usepackage{algpseudocode}
\usepackage[compress]{cite}
\usepackage{amsmath,amsfonts}
\usepackage{listings}
\usepackage{mathtools}
\usepackage{multirow}
\usepackage{subcaption}
\usepackage{array}
\usepackage{footnote}
\usepackage{kotex}

\def\figref#1{Fig.~\ref{#1}}

\def\secref#1{Section~\ref{#1}}

\def\tabref#1{Table~\ref{#1}}

\long\def\ignore#1{}

\def\bottomoffigure{}  
\newcommand{\myincludegraphics}[3][]{\centering
\includegraphics[#1]{#2}\caption{#3\label{#2}}\bottomoffigure}
\graphicspath{{figs/}}

\usepackage[backgroundcolor=green!20]{todonotes}
\newcounter{todocounter}
\makeatletter\@ifpackageloaded{todonotes}{\newcommand{\todonum}[2][]{\stepcounter{todocounter}\todo[inline]{\thetodocounter: #2\ifx\x#1\x{}\else{ [#1]}\fi}}}{\newcommand{\todonum}[2][]{\par\stepcounter{todocounter}\noindent\fcolorbox{black}{green!20}{\begin{minipage}{.98\linewidth}\thetodocounter: #2\ifx\x#1\x{}\else{ [#1]}\fi\end{minipage}}}}\makeatother

\makeatletter\@ifclassloaded{IEEEtran}
{\newenvironment{keyword}{\begin{IEEEkeywords}}{\end{IEEEkeywords}}
\ifCLASSOPTIONconference\fi}{
\def\bstctlcite{\@ifnextchar[{\@bstctlcite}{\@bstctlcite[@auxout]}}
\def\@bstctlcite[#1]#2{\@bsphack\@for\@citeb:=#2\do{%
\edef\@citeb{\expandafter\@firstofone\@citeb}%
\if@filesw\immediate\write\csname #1\endcsname{\string\citation{\@citeb}}\fi}%
\@esphack}}\makeatother

\begin{document}

\title{NTT-PIM: Row-Centric Architecture and Mapping for Efficient Number-Theoretic Transform on PIM%
\thanks{This work was supported by the Samsung Advanced Institute of Technology, Samsung Electronics Co., Ltd., by IITP grant (No.~2020-0-01336, Artificial Intelligence Graduate School Program (UNIST)), and by the Free Innovative Research Fund of UNIST (1.170067.01).

\textsuperscript{\textsection}J.~Lee is the corresponding author (E-mail: jlee@unist.ac.kr).}}

\ignore{
\author{\IEEEauthorblockN{Jaewoo Park}
\IEEEauthorblockA{Department of Physics, UNIST \\
Ulsan, South Korea\\
hecate64@unist.ac.kr}
\and
\IEEEauthorblockN{Sugil Lee}
\IEEEauthorblockA{Department of EE, UNIST \\
Ulsan, South Korea\\
sglee17@unist.ac.kr}
\and
\IEEEauthorblockN{Jongeun Lee}
\IEEEauthorblockA{Department of EE, UNIST \\
Ulsan, Korea\\
jlee@unist.ac.kr}
}}

\author{%
  \IEEEauthorblockN{%
    Jaewoo Park\IEEEauthorrefmark{1},
    Sugil Lee\IEEEauthorrefmark{2} and
    Jongeun Lee\IEEEauthorrefmark{2}\textsuperscript{\textsection} 
  }%
 \IEEEauthorblockA{\IEEEauthorrefmark{1}Department of Physics,
 \IEEEauthorrefmark{2}Department of Electrical Engineering}%
\IEEEauthorblockA{Ulsan National Institute of Science and Technology (UNIST), Ulsan, Korea}%
\IEEEauthorblockA{\texttt{\{hecate64,sglee17,jlee\}@unist.ac.kr}}%
}

\maketitle


\bstctlcite{IEEEexample:BSTcontrol}

\begin{abstract}
\ignore{
Recently digital PIM (Processing-in-Memory) has received renewed interest for memory intensive applications such as AI.
For non-AI applications, however, NTT (Number-Theoretic Transform) for example is the most important kernel for FHE (Fully Homomorphic Encryption) algorithms but its drastically different characteristics than AI applications makes it difficult to accelerate with AI-PIM.
In this paper we propose a novel PIM architecture that supports more flexible data movement within and across bank rows, and an efficient mapping and scheduling method for accelerating applications having non-sequential memory accesses such as NTT.
}
Recently DRAM-based PIMs (processing-in-memories) with unmodified cell arrays have demonstrated impressive performance for accelerating AI applications.  However, due to the very restrictive hardware constraints, PIM remains an accelerator for simple functions only.  In this paper we propose NTT-PIM, which is based on the same principles such as no modification of cell arrays and very restrictive area budget, but shows state-of-the-art performance for a very complex application such as NTT, thanks to features optimized for the application's characteristics, such as in-place update and pipelining via multiple buffers. 
Our experimental results demonstrate that our NTT-PIM can outperform previous best PIM-based NTT accelerators in terms of runtime by $1.7\sim 17\times$ while having negligible area and power overhead. 
\end{abstract}


\section{Introduction}
The recent success of AI has caused important changes in the computing landscape, one of which being the renewed interest in PIM (processing-in-memory) \cite{eet,fim,9489313}.  Though the idea of PIM dates back to 70's, recent approaches to AI-PIM by DRAM makers take a more principled approach \cite{newton-isscc22}, such as limiting the size of compute hardware, providing a full SW stack, and keeping memory arrays and the DRAM interface intact, which has led to successful demonstrations of AI-PIM in the industry setting  \cite{newton-isscc22,kwon2021fimdramm}.

On the other hand, current AI-PIM architectures \cite{mert2022anextensive,he2020newton} can only support very simple functions such as matrix-vector multiplication (MVM).
In this paper we extend the scope of DRAM-based PIM by optimizing an industry-designed PIM architecture \cite{he2020newton} for a nontrivial non-AI application.  In particular, we target Fully Homomorphic Encryption (FHE) \cite{10.1007/978-3-319-70694-8_15}, where the most important function is NTT (Number-Theoretic Transform).  In addition to being memory-bound, NTT has highly irregular memory access patterns, which is a main difference compared to MVM or AI applications.


In this paper we propose a novel PIM architecture and a mapping method for NTT on PIM.
One of the key challenges in finding efficient mapping is asymmetric memory access time depending on whether consecutive accesses to a bank are to the same row (\emph{buffer hit}) vs.\ different rows (\emph{buffer conflict}).
While previous works on accelerating FHE workloads \cite{riazi2020heax, samardzic2021f1} use large on-chip memory to increase data reuse and hide memory latency, PIM must economize on chip area, which is another key challenge.
Exploiting the recursive structure of the NTT computation, our mapping algorithm divides the problem into smaller ones, and uses a different mapping strategy depending on the input size, categorized into three regimes based on architectural parameters.
One important architectural parameter is the size and number of local buffers.  We show that whereas a single-buffer architecture is extremely inefficient for NTT, providing at least one auxiliary buffer would greatly enhance mapping efficiency through a technique called \emph{in-place update}.
We also propose and evaluate the use of more buffers through \emph{pipelining}, which increases hardware overhead minimally but can improve performance significantly by both hiding memory latency and reducing the number of row activations.
Our architecture and mapping scheme also allows for bank-level parallelism (i.e., running different NTT functions in each bank), which can be naturally exploited by FHE applications with linear speedup. \ignore{(e.g., 16$\times$ over single-bank mapping on HBM (High-Bandwidth Memory)).}

Our experimental results based on architectural parameters of HBM2E demonstrate that our PIM architecture has very little hardware overhead, less than half of Newton's \cite{he2020newton}, which is already at a tiny level, yet can deliver state-of-the-art performance for NTT acceleration.  Compared with the previous best PIM-based NTT accelerators \cite{li2022mentt,nejatollahi2020cryptopim}, our architecture does not modify cell arrays, and our solution can support arbitrary polynomial length and modulo values, and yet ours can deliver up to $1.7\sim 17\times$ speedup at the NTT level (except the bit reversal, which is common in all the compared works).

\ignore{
(what about cost or energy?).  

can deliver state-of-the-art performance at very little 
 outperforms latest PIM based NTT accelerators by $2\sim 3\times$ speedup on computation latency and $7\sim 17\times$ compared to CPU  just by using a single BU on the DRAM memory die.}

\ignore{In this paper we make the following contributions.
\begin{itemize}
\item Novel PIM architecture for NTT, including auxiliary atom buffer(s). 
\item mapping algorithm tailored for PIM hardware constraint (?)
\item Efficient mapping method exploiting in-place update and pipelining, which can simultaneously save hardware cost and minimize memory latency.
\item Evaluation.
\end{itemize}
}

\section{Background and Related Work}

\begin{figure}
  \myincludegraphics[width=.95\linewidth]{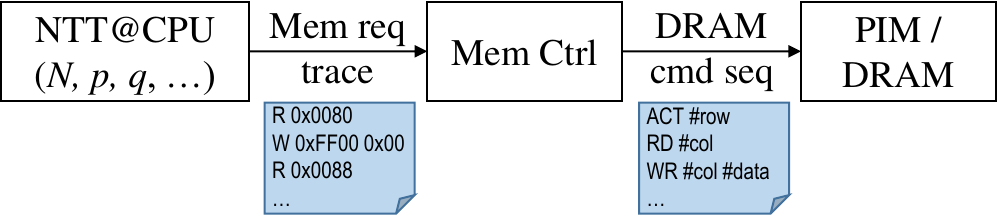}{From software to DRAM commands.}
\end{figure}

\begin{figure}
  \myincludegraphics[width=.95\linewidth]{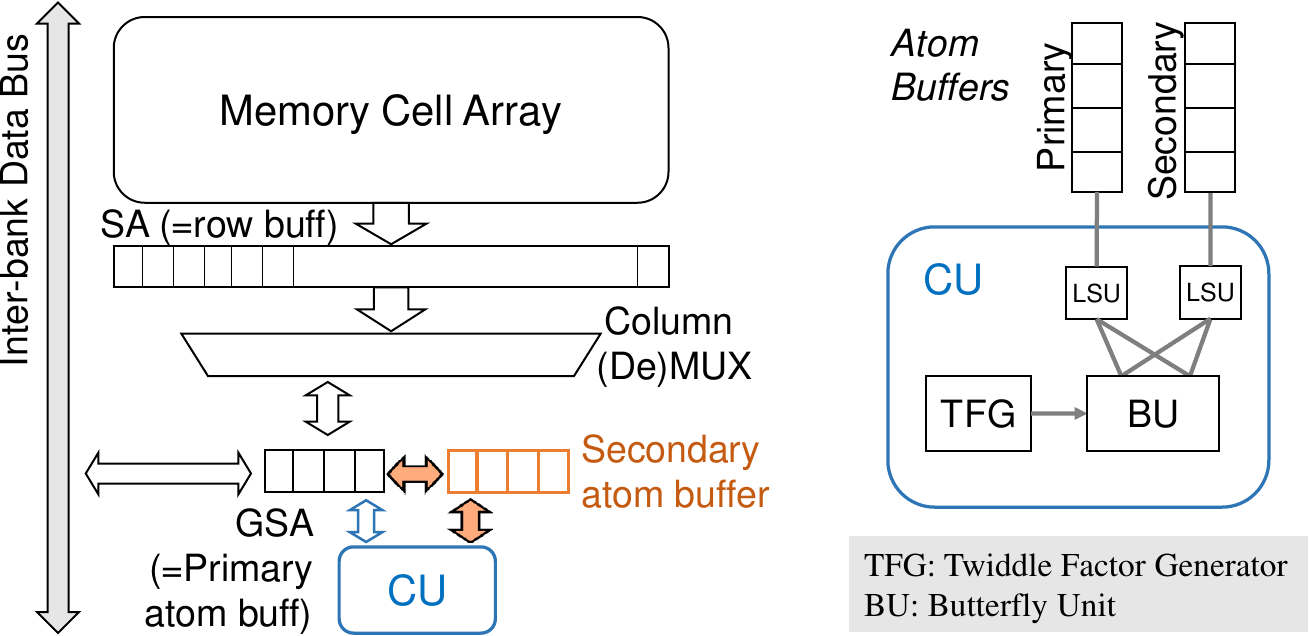}{Left: Bank datapath, where colored parts are our extension.  Right: Compute Unit (CU).}
\end{figure}

\subsection{DRAM and PIM}

Software running on a CPU issues memory requests, which are handled by the memory controller (MC) (see \figref{flow}).  Understanding DRAM timing, MC schedules memory requests and generates low-level DRAM commands such as row activate and column read/write.
A DRAM chip consists of multiple banks, which share command, address, and data buses.  At the chip and bank level, the unit of a memory transaction is a DRAM \emph{atom} (32B in HBM), which ultimately comes from a row of DRAM cells of a bank \cite{chatterjee2017hpca}.
When a row is activated, the contents of the cells belonging to the row are copied to the bitline sense amplifiers (BLSAs), also known as \emph{row buffer} (e.g., 1 KB in size).  A subsequent read command latches part of the row buffer (a DRAM atom) to the global sense amplifiers (GSAs), sending it out via chip I/O (see \figref{arch}). 

Recent PIM works add extra logic inside DRAM banks \cite{kwon2021fimdramm} or modify the dram die structure \cite{newton-isscc22}. DRAM based PIM has been able to successfully integrate computing logic inside existing DRAM technology, significantly reducing the memory bottleneck of machine learning applications. While the processing unit of FIMDRAM \cite{kwon2021fimdramm} is located in an independent memory cell and uses an additional bus to send data among banks, Newton \cite{newton-isscc22} has a MVM logic inside the memory bank, completely eliminating the data transfer between the logic and the memory. The limitation of Newton \cite{newton-isscc22} is that the datapath is rather simple and fixed, and it is hard to store intermediate data. This constraint makes it hard to map algorithms with intermediate data and complex memory access such as NTT.

MeNTT \cite{li2022mentt} is a 6T-SRAM based PIM accelerator for NTT. It computes NTT in a bit-serial fashion, which is not very efficient for high-precision arithmetic such as 64-bit or 32-bit.  Also it is inflexible in terms of modulo and maximum polynomial length.
CryptoPIM \cite{nejatollahi2020cryptopim} is a ReRAM-based PIM accelerator targeting polynomial multiplication. While it is expected to have low cost due to the advantage of ReRAM technology, ReRAM technology has issues with fabrication such as variation and faults.
Also CryptoPIM is inflexible in terms of modulo and maximum polynomial length.


\subsection{FHE and NTT}
Latest FHE schemes \cite{cryptoeprint:2012/144,10.1007/978-3-319-70694-8_15} are based on Ring Learning-With-Errors (Ring-LWE) \cite{10.1007/978-3-642-13190-5_1}, which encodes vectors as polynomials with coefficients of finite fields.  A typical polynomial can be defined as $R_q = \mathbb{Z}_q[X]/(X^N+1)$, a polynomial whose length is a power of 2 and each of its coefficients is a modulo of a prime number $q$. The product of two polynomials can be computed efficiently in the NTT domain \cite{Pppelmann2012TowardsEA}.  NTT is a generalization of the discrete Fourier transform (DFT) on the finite field, where complex multiplication is replaced with an integer multiplication followed by a modulo operation.  For a given polynomial $a = a_0 + a_1x + a_2x^2 + ... + a_{N-1}x^{N-1}$, we can associate a vector $\mathbf{a} = (a_0, a_1, ... a_{N-1})$.
Then the polynomial multiplication $a*b$ can be computed using NTT, which is more efficient than direct multiplication.
\begin{equation}
    \mathbf{a*b} = NTT^{-1}(NTT(\mathbf{a}) \odot NTT(\mathbf{b}))
\end{equation}
where $\odot$ represents the element-wise product and $NTT^{-1}$ is inverse NTT.  The inverse NTT computation is mathematically identical to the original NTT except that the twiddle factor $\omega$ is replaced with its inverse.

\begin{figure}
\centering
\begin{subfigure}{.78\linewidth}
  \includegraphics[width=\linewidth]{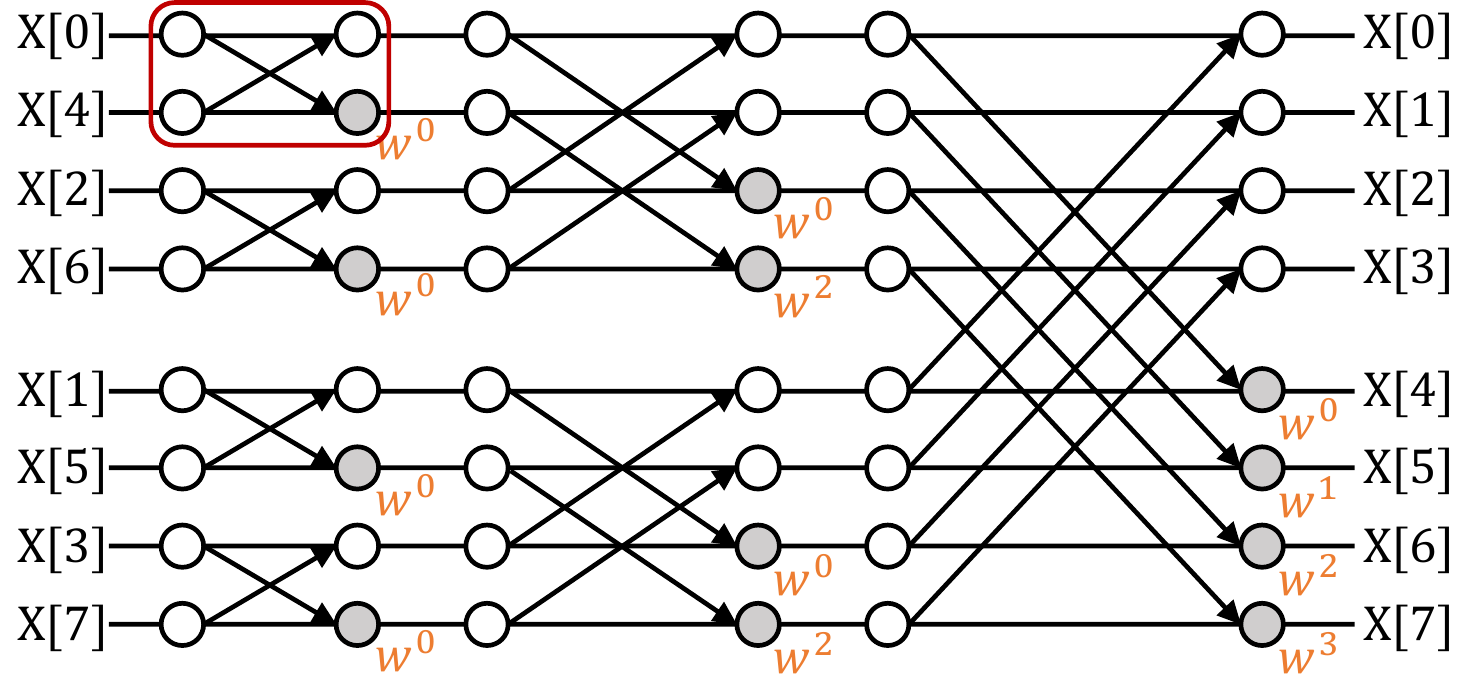}
\end{subfigure}
\hfill
\begin{subfigure}{.20\linewidth}
  \includegraphics[width=\linewidth]{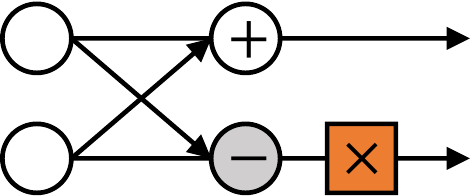}
\end{subfigure}
\caption{Left: Dataflow of NTT algorithm (based on Cooley-Tukey FFT, bit-reversal is omitted).
  Right: A butterfly unit (BU) operation consisting of two ModAdd/Sub and one ModMult operations.}
\label{ntt-ct8}
\end{figure}

NTT shares the characteristic irregular memory access pattern with FFT (Fast Fourier Transform).  A size-$N$ NTT can be computed in $\log N$ stages, each of which involves $N/2$ parallel butterfly unit (BU) operations, followed by bit reversal (see \figref{ntt-ct8}).
There is a large body of work on efficient parallel FFT algorithms \cite{10.1145/321450.321457, 10.1016/S0167-8191(84)90413-7},
e.g., Pease \cite{10.1145/321450.321457}, which is well suited for FPGAs and ASICs due to its regular structure, and Stockham \cite{10.1016/S0167-8191(84)90413-7}, which is self-sorting.
However, those algorithms often require multiple ($\log N$) shuffling stages, which means more frequent interactions with CPU (unless hardware for shuffling is added, increasing cost)
and may not be any better than the Cooley-Tukey FFT algorithm \cite{Harris1977VectorRF}.  Also intra-row data reuse can be easily and fully exploited by recursive Cooley-Tukey FFT, which we use in this work.  We assume that bit reversal is performed by software running on a CPU, which is a common assumption in previous PIM approaches \cite{li2022mentt,nejatollahi2020cryptopim}.  Further, bit reversal can be avoided altogether when all NTT-domain operations are element-wise operations \cite{li2022mentt}.


\begin{figure}
  \myincludegraphics[width=.9\linewidth]{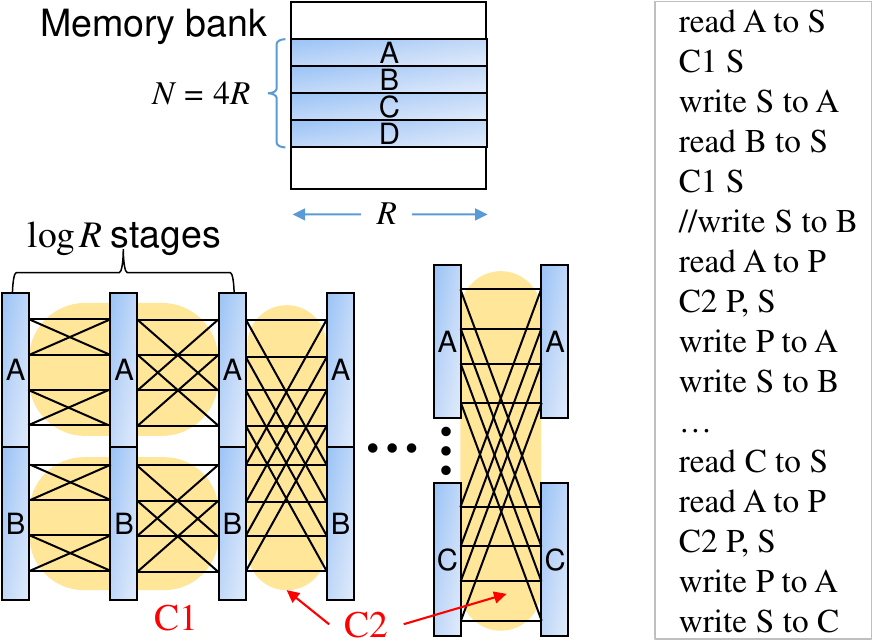}{Mapping example ($P$ and $S$ represent local buffers).}
\end{figure}

\section{Our Architecture and Mapping}

\subsection{Analysis of NTT Computation}

For size-$N$ NTT, the number of compute operations is $O(N \log N)$, and the amount of data transfer is either $O(N)$ if the entire input fits in the local memory, whose size is denoted by $M$, or $O(N)$ \emph{per stage} otherwise.  In recursive Cooley-Tukey FFT, the first $\log M$ stages (assuming $M$ is a power of 2) can be partitioned into $\frac{N}{M}$ independent blocks, each of which would fit in the local memory (see \figref{mapex1}).  Therefore, the total amount of data transfer when $N \ge M$ is $O(N + N (\log N - \log M))$, and 
the compute-to-data-transfer ratio (CDR) is 
 $O(\frac{\log N}{1 + \log \frac{N}{M}}) \le O(\log N)$, where equality holds when $M = N$.
The CDR cannot be increased further by such techniques as batching.   Tiling (also known as blocking) is also difficult due to the irregular memory access pattern.
All of the above suggests that NTT is memory-bound and PIM can be a good candidate to accelerate NTT.
However, DRAM memories have very asymmetric access time.
Also the very limited size of a row compared to $N$ makes the problem of mapping NTT to PIM challenging.



\subsection{Necessity of An Auxiliary Buffer}
Suppose $N$ is small enough to fit in a row buffer.  Then the NTT function can be implemented efficiently by a small compute unit (CU) within a bank that can read two words of a row (performed by two \textbf{load} $\mu$-ops), do a butterfly unit (BU) operation, and write the result back to the input locations (by two \textbf{store} $\mu$-ops).  Every access to memory (i.e., load/store) except the first will be a buffer hit, which is optimal. 


However, if $N$ is larger than the row buffer size (denoted by $R$), there is no efficient mapping that realizes maximal data reuse, unless there is an extra buffer.  With a single buffer (i.e., GSA only), assuming CU is equipped with two scalar registers for two BU inputs (see \figref{arch}),
\emph{every BU operation} requires two loads and two stores, and about half of them require row activation, which greatly increases memory access time.

\subsection{Sufficiency of One Auxiliary Buffer}
Suppose that $N = 4R$.  As illustrated in \figref{mapex1},
the first $\log R$ stages can be vertically partitioned into 4 independent blocks (only two are shown), each of which can be processed efficiently with one row activation, resulting in just 4 ($= \frac{N}{R}$) row activations for the $\log R$ stages.

The question is whether the other stages can be done efficiently.
Let us assume that there is an auxiliary buffer of size $R$.  We call GSAs and the extra buffer \textit{primary} ($P$) and \textit{secondary atom buffer} ($S$), respectively. 
For the $(1 + \log R)$-th stage, a na\"ive implementation may require a third buffer for the output, since the two buffers $P$ and $S$ are occupied by the two inputs.
But observing that each input element is used exactly twice, we can schedule operations such that we complete each BU operation before moving on to the next, which allows us to store the output of a BU operation directly to its input locations (\emph{in-place update}).
The combination of BU-grained scheduling and in-place update not only eliminates the need for a separate output buffer,
but also allows us to fully reuse input data in $P$ and $S$, which is true for the following stages as well.
Each \textit{stage} (not a BU op) after the first $\log R$ stages requires 4 ($= \frac{N}{R}$) pairs of reads and writes, but half of the writes can be made a buffer hit, resulting in at most $3\frac{N}{2R}$ row activations, which is much less than if no in-place update and no third buffer are used.

\subsection{New DRAM Commands and Our PIM Architecture}
One problem with adding an extra row buffer is a very large area overhead.  However, the extra buffer needs only to be of a DRAM atom size, $N_a$.  In fact, GSAs are also of a DRAM atom size \cite{pum16,chatterjee2017hpca}.  

To realize the mapping outlined above, CU needs to support the following operations.
\begin{itemize}
\item \textbf{C1} takes input from one buffer, $S$, and performs the NTT function minus bit-reversal, which includes $\log N_a$ stages of $\frac{N_a}{2}$ BU operations per stage.  The result is stored into the same buffer as the input.
\item \textbf{C2} takes input from two buffers, $P$ and $S$, and performs one $N_a$-way vectorized BU operation.  The result is stored back into $P$ and $S$.
Note that C1 and C2 must perform modulo add/mult \cite{mert2022anextensive} in order to support NTT.
\end{itemize}


The memory hierarchy of our PIM architecture is as follows.  The CU has two operand registers (as well as other registers, e.g., for twiddle factors), which can be seen as L0 memory.  The $P$ and $S$ buffers are L1 memory, and data movement between L0 and L1 is accomplished by load/store $\mu$-ops, which are very fast (2 cycles) and part of C1 and C2.  The memory banks are L2 memory, and data transfer between L1 and L2 is managed by explicit commands, which we simply call \textbf{read} and \textbf{write}.  These CU-read/write are similar to column read/write, taking about a dozen cycles, except that data transfer stops at $P$ or $S$ instead of chip I/O.

\begin{algorithm} 
  \caption{Intra-atom NTT compute: $\mathrm{C1}(S) \to S$}
  \label{alg-c1}
  \textbf{Inout} $S$ : vector reg for both input and output\\
  \textbf{Param} $\omega_0, r_{\omega}$ : init value \& step size for twiddle factor gen.
  \begin{algorithmic}[1]
    \State $\omega_s$ : twiddle factor for each stage, initialized to $1$
    \For {$(stage=1; stage \leq \log{N_a}; stage\,\texttt{+=}1)$}
      \State \textbf{let} $\omega \leftarrow \omega_0 $, $m \leftarrow 2^{stage-1}$
      \For {$(k=0; k < N_a; k \,\texttt{+=}2m)$}
            \For {$(j=0; j < m; j \,\texttt{+=}1)$}
            \State \textbf{let} $a \leftarrow S[k+j]$
            \State \textbf{let} $b \leftarrow S[k+j+m]$
            \State $S[k+j] \leftarrow (a+b) \mod q$
            \State $S[k+j+m] \leftarrow (a-b)\cdot \omega \mod q$
            \State  $\omega \leftarrow  \omega \cdot \omega_{s} \mod q$
        \EndFor
      \EndFor
      \State  $\omega_{s} \leftarrow  \omega_s \cdot r_{\omega} \mod q$
    \EndFor
  \end{algorithmic} 
\end{algorithm}

\ignore{
\begin{algorithm} 
	\caption{Na\"ive Inter-row NTT ($M$, $r$, $w$) $\to$ ??} 
        \textbf{Input} $M$ : the size of NTT poly (must be powers of 2) \\
        \textbf{Input} $r$ : the row offset for the NTT poly \\
        \textbf{Input} $w$ : the first twiddle factor 
	\begin{algorithmic}[1]
        \State $\omega_0 \leftarrow \omega $
	\For {$(stage=1; stage \leq \log_2{N_a}; stage \mathrel{+}=1)$}
            \State loadbuf($0, r+2^s$)
                \State $\omega \leftarrow \omega_0 $
			\For {$k=0,1,\ldots,N/2$}
                    \State $b1[k] \leftarrow b0[k*2^{log(N/2)-s}] + w*b0[k*2^{log(N/2)-s}+2^{s}]$
                    \State  $\omega_0 \leftarrow  \omega_0 * \omega_{s} $
			\EndFor
        
                \State  $\omega_{s} \leftarrow  \omega_{s} * \omega $
            \State storebuf($1, r+2^s$)
        \EndFor
	\end{algorithmic} 
\end{algorithm}
}

\begin{algorithm}
    \caption{Inter-atom NTT compute: $\mathrm{C2}(P, S) \to P, S$}
    \label{alg-c2}
    \textbf{Inout} $P, S$ : vector regs for both input and output\\
    \textbf{Param} $\omega_0, r_{\omega}$ : init value \& step size for twiddle factor gen.
    \begin{algorithmic}[1]
    \State \textbf{let} $\omega \leftarrow \omega_0 $
    \For {$(j=0; j < N_a;  j \,\texttt{+=}1)$}
            \State \textbf{let} $a \leftarrow P[j]$, $b \leftarrow S[j]$
            \State $P[j] \leftarrow (a+b) \mod q$
            \State $S[j] \leftarrow (a-b) \cdot \omega \mod q$
            \State  $\omega \leftarrow  \omega \cdot r_{\omega} \mod q$
        \EndFor
	\end{algorithmic} 
\end{algorithm}

\section{Implementation Detail}

\subsection{Host Interface and Architecture}

From the software perspective, our NTT function can be invoked as a write request (see \figref{flow}), which contains NTT parameters as ``write data''.  We assume $N$ to be a power of 2.  The input data ($N$ 32b integers) is assumed to be already in the memory; thus, only the address is passed.  The MC generates a sequence of DRAM commands to fulfill the NTT function (see \figref{mapex1}).  The result is stored at the same location as the input, and a write response is given to the request initiator to signal that NTT is completed.
The MC needs to be modified to implement the mapping algorithm presented in the next section.

Algorithms~\ref{alg-c1} and \ref{alg-c2} describe the C1 and C2 commands.
The BU in \figref{arch} has two operand registers.  Each buffer is single-ported, but a small crossbar switch allows full connectivity between buffers and BU registers. 
To do computation, CU first loads input from $P$ and/or $S$ to its registers, does a BU operation, and stores the results, which is repeated in a pipelined fashion for all elements available.  In our architecture the size of a DRAM atom ($N_a$) is 8.

We generate twiddle factors on-the-fly using a method similar to \cite{6581570}, which allows us to use all memory bandwidth for accessing input polynomial.


To pass parameters ($q$, $\omega_0$, $r_{\omega}$), a value (16-bit) is placed on the global buffer, which is visible to all banks, and subsequently loaded to a scalar register in the CU of a bank (in multiple cycles for higher precision values). 

The secondary atom buffer can be implemented using SRAM cells (6T/cell) plus inverters (2T/cell) to provide complementary signals needed. 


\begin{figure*}
  \def\xx{13.5mm}
\begin{subfigure}{.2\linewidth}
  \myincludegraphics[height=\xx]{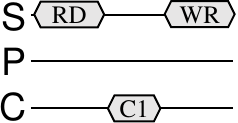}{Intra-atom mapping}
\end{subfigure}
\begin{subfigure}{.37\linewidth}
  \myincludegraphics[height=\xx]{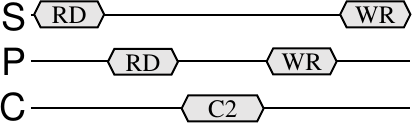}{Intra-row mapping}
\end{subfigure}
\begin{subfigure}{.45\linewidth}
  \myincludegraphics[height=\xx]{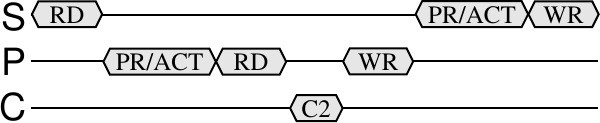}{Inter-row mapping}
\end{subfigure}
\caption{Timing diagram for three mapping regimes (S/P: secondary/primary atom buffer, C: CU). Initial row activate is omitted.}
\label{timing}
\end{figure*}

\ignore{
\subsection{DSP-PIM Architecture}

\begin{figure}
  \myincludegraphics[width=.3\linewidth]{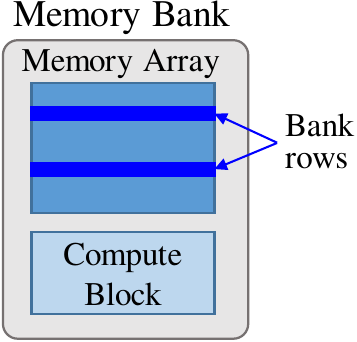}{Memory bank.}
\end{figure}

1. Butterfly Unit

2. Extra row buffer (works like a cache)

3. LU/SU hardware (Butterfly Access pattern 지원하기 위해)

}

\subsection{NTT Computation Mapping}

Given the memory hierarchy of our PIM architecture, there are three mapping regimes as illustrated in \figref{timing}. 
\begin{enumerate}
\item Intra-atom (when $N \le N_a$): applies to the first $\log N_a$ stages, and uses C1 command.
\item Intra-row (when $N_a < N \le R$, where $R$ is the row buffer size): applies to the next $\log \frac{R}{N_a}$ stages.
\item Inter-row (when $N > R$): applies to the rest of the stages.  Both intra-row and inter-row use C2 command.
\end{enumerate}
Both intra-row and inter-row are ``inter-atom'', but their difference is that only inter-row requires intermittent row activate commands.

The MC generates DRAM commands by dividing a given NTT function or its dataflow graph (DFG) such as \figref{ntt-ct8} as follows
\ignore{(assuming the input array is aligned to $R$-words boundary)}.  First, note that there are two ways to divide a DFG: stage-wise (horizontally) or data-wise (vertically).  Second, observe that all data dependence is within a row during the first $\log {R}$ stages, whereas in the latter stages all data dependence is across rows but BU operations are vectorized with at least $R$ ways.  (The analogous applies to an atom buffer as well.)  Therefore, we propose to divide the first $\log {R}$ stages of a DFG vertically into $\frac{N}{R}$ independent row-sized blocks, each of which is divided again horizontally into (i) the first $\log N_a$ stages, which are handled using intra-atom mappding, and (ii) the rest of the stages, which are handled using intra-row mapping. 
The latter stages are in the inter-row regime, for which we process a DFG stage-by-stage and each stage in the sequential order.  Since each C2 command in this regime always involves at least two row activations, the processing order does not make much difference in performance, but increasing the size or number of buffers does, which we discuss next. 

\ignore{
\begin{algorithm}
  \caption{Mapping algorithm for a single row-sized block}
\label{alg-intra-row}
\begin{algorithmic}[1]
\State \texttt{activate} row
\For {subblock $i$ \textbf{in} $1, \cdots, \frac{R}{N_a}$}
  \State \texttt{intra\_atom} for subblock $i$    (handles $\log {N_a}$ stages)
\EndFor
\For {stage $s$ \textbf{in} $1 + \log {N_a}, \cdots, \log {R}$}
   \State // each stage has $\frac{R}{N_a}$ atom-sized blocks
   \For {subblock $i$, $j$ \textbf{in} $1, \cdots, \frac{R}{2N_a}$}
      \State \texttt{intra\_row} for stage $s$, subblock $i$
   \EndFor
\EndFor
\State \texttt{precharge} row
\end{algorithmic}
\end{algorithm}
}

\ignore{
First, the first $\log N_r$ stages are divided into $\frac{N}{N_r}$ blocks of independent intra-row computation.  For each intra-row block, 
First, the first $\log N_a$ stages are divided into independent intra-atom blocks, the code for which is generated and executed sequentially, then (ii) 
the next $\log \frac{N_r}{N_a}$ stages are divided into $\frac{N}{N_r}$ blocks of independent intra-row blocks. 

which comprises the first 3 (= $\log N_a$) stages.
}

\section{Pipelining Optimization}

\begin{figure*}
  \def\xx{20.0mm}
\begin{subfigure}{.47\linewidth}
  \myincludegraphics[height=\xx]{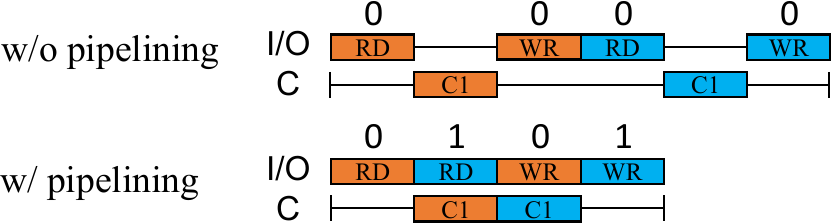}{Intra-atom mapping ($N_b$ = 1 vs.\ 2)}
\end{subfigure}
\begin{subfigure}{.45\linewidth}
  \myincludegraphics[height=\xx]{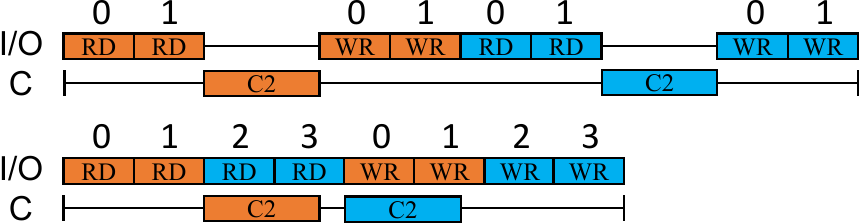}{Intra-row mapping ($N_b$ = 2 vs.\ 4)}
\end{subfigure}
\begin{subfigure}{.9\linewidth}
\vspace{4mm}
  \myincludegraphics[height=\xx]{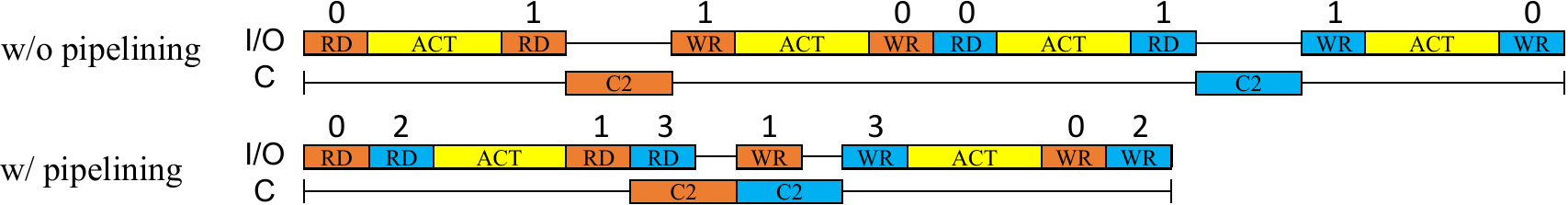}{Inter-row mapping ($N_b$ = 2 vs.\ 4)}
\end{subfigure}
\caption{When two CU operations are executed without vs.\ with pipelining.  Same color means dependence, numbers buffer indices, and $N_b$ is the number of buffers (incl. primary).  Performance improvement is due to (i) overlapping memory latency with compute latency, and in the case of inter-row mapping, also to (ii) reducing the number of row activations.}
\label{pipeline}
\end{figure*}

While a dual-buffer architecture can fully realize data reuse up to intra-row mapping, it suffers from frequent row activations in inter-row mapping.
Also even if full data reuse is achieved, it does not necessarily mean the highest throughput possible.
\figref{pipeline} illustrates the difference between executing two CU operations consecutively (without pipelining) vs.\ with pipelining.  In the case of pipelined execution, read commands for the second compute operation can start before write commands for the first compute finish, thereby hiding some memory access latency.  

On the other hand, pipelining requires more buffers.  In the case of intra-atom mapping, pipelining is possible even with a single auxiliary buffer (since our mapping uses only one buffer, we can use the other buffer for pipelining), but more buffers mean more overlap.  In general, to overlap $n$ executions requires $n$ times as many buffers.  For the other mapping regimes, more than one auxiliary buffers are necessary for pipelining.  Also C1 and C2 commands need to take operands from any buffers, including one primary and multiple secondary buffers.  Thus in addition to increased buffer area, hardware overhead of pipelining includes a larger crossbar switch in CU (see \figref{arch}).  CU-read/write commands and C1/C2 commands need more parameters too, but having no immediate field, they have enough bits to encode additional parameters. 

Pipelining has an unexpected positive effect in inter-row mapping: reduced number of row activations. 
Equipped with more buffers thanks to pipelining, we can group same-row read/write commands together as illustrated in \figref{pipeline-c}, eliminating some row activations in the process.  Note that the likelihood of finding same-row commands within the two CU compute operations is quite high, since the inter-row regime has highly vectorized BU operations. 

The MC is responsible for generating pipelined schedules.  One issue is that since the command bus is shared between all DRAM commands, we need to choose which command to issue first between read/write vs.\ compute.  It can be decided rather easily, since the latency of each command is known in advance.  

\section{Experiments}

\subsection{Experimental Setup}

To evaluate the performance of our NTT-PIM, we have developed our in-house PIM simulator, which consists of a front-end driver and DRAMsim3 \cite{dramsim3} working in tandem. 
The front-end driver serves two purposes: to generate DRAM command sequence simulating the memory controller including our mapping algorithm, and to verify the functionality of our NTT function as executed in DRAMsim3.\footnote{The front-end driver is written in Python, and runs iteratively with DRAMsim3 to implement two-way communication between the two, which was added to double-check the correctness of timing and functionality.}

\tabref{t-param} lists the parameters used in our evaluation, which are based on HBM2E.  
While all our results are based on running NTT on a single bank, FHE applications can naturally run multiple NTT functions using multiple banks. 

\begin{table}[]
\caption{Architecture Parameters} \label{t-param}
\centering
\begin{tabular}{|lr||lr|}
\hline
\multicolumn{2}{|c||}{Architecture Parameters} & \multicolumn{2}{c|}{Timing Param. (cycles)} \\ \hline
DRAM atom size                & 32B     & CL                    & 14                 \\
\# of columns per row    & 32      & tCCD                  & 2                  \\
\# of rows per bank      & 32,768   & tRP                   & 14                 \\
\# of ranks          & 1       & tRAS                  & 34                 \\
\# of banks          & 1       & tRCD                  & 14                 \\
                         &         & tWR                   & 16                 \\ \hline
\end{tabular}
\end{table}

\subsection{Synthesis Results}

Since PIM is supposed to be fabricated using a memory technology, which is not open, it is hard to estimate the area overhead of our architecture accurately.  Instead, we provide a comparison with Newton \cite{he2020newton} in terms of hardware overhead, which should assure that our CU is small enough to be fabricated inside a memory bank. 
We have implemented both Newton's compute hardware, mostly consisting of 16 16-bit floating-point MACs, and our CU in \figref{arch} in Verilog RTL, and synthesized them with Synopsys Design Compiler using Samsung 65 nm standard-cell library.
The area of auxiliary atom buffers is estimated with CACTI 7.0 at 65 nm.
We have verified that our designs meet the timing requirement for 1200 MHz, which is the operating frequency of HBM2E.

We have fully pipelined the BU inside CU, which supports ModAdd/Sub and ModMult for arbitrary modulo values using Montgomery reduction algorithm \cite{Montgomery1985ModularMW}.  The latency of C1 and C2 is 15 and 10 cycles, respectively.

\begin{table}
\centering
\caption{PIM Area Overhead ($N_b$: \# of all atom buffers)}
\begin{tabular}{|cc|rc|}
\hline
Architecture             &  $N_b$  & Area (mm$^2$) & \%   \\ \hline
A DRAM bank              &     & 4.2208               & --     \\ \hline
Newton \cite{he2020newton} &   & 0.0474                & 1.123 \\ \hline
\multirow{4}{*}{NTT-PIM} &  1  & 0.0213                & 0.504 \\
                         &  2  & 0.0232                & 0.550 \\
                         &  4  & 0.0263                & 0.624 \\
                         &  6  & 0.0285                & 0.676 \\ \hline
\end{tabular}
\label{t-area}
\end{table}

\tabref{t-area} summarizes area results, which show that our hardware overhead is small, being less than half of Newton's, which is already at a tiny level.  Also the additional overhead of having multiple atom buffers seems marginal.
To give a rough comparison, we have also estimated the area of a single DRAM bank using CACTI-3DD \cite{cacti-3dd} DDR4 model at 32~nm, which is the most advanced node supported by the tool.\footnote{We intentionally use an older technology for logic, to simulate the effect of implementing logic with a memory technology, which would be slower and take larger area.}



\begin{table*}[ht]
\centering
\caption{Comparison with Previous Work}
\label{t-cmp}
\begin{tabular}{cr|rrrrrrr}
\hline
\multicolumn{2}{c|}{Design}                               & \multicolumn{3}{c}{NTT-PIM}                                           & \multicolumn{1}{c}{MeNTT}                    & \multicolumn{1}{c}{CryptoPIM}                & \multicolumn{1}{c}{x86 CPU}             & \multicolumn{1}{c}{FPGA}                \\ \hline
\multicolumn{2}{c|}{Method}                               & \multicolumn{3}{c}{DRAM}                                              & \multicolumn{1}{c}{6T-SRAM}                  & \multicolumn{1}{c}{RRAM}                     & \multicolumn{1}{c}{Software}            & \multicolumn{1}{c}{-}                   \\ \hline
\multicolumn{2}{c|}{Bitwidth}                             & \multicolumn{3}{c}{32}                                                & \multicolumn{1}{c}{\multirow{2}{*}{14 / 16}} & \multicolumn{1}{c}{\multirow{2}{*}{16 / 32}} & \multicolumn{1}{c}{\multirow{2}{*}{32}} & \multicolumn{1}{c}{\multirow{2}{*}{16}} \\
\multicolumn{2}{c|}{\# of atom buffers ($N_b$) }                  & \multicolumn{1}{c}{2} & \multicolumn{1}{c}{4} & \multicolumn{1}{c}{6} & \multicolumn{1}{c}{}                         & \multicolumn{1}{c}{}                         & \multicolumn{1}{c}{}                    & \multicolumn{1}{c}{}                    \\ \hline
\multicolumn{1}{c|}{\multirow{5}{*}{Latency (ns)}} & $N$: 256  & 3.90                  & 2.50                  & 1.94                  & 23$^\ddagger$~~                                         & 68.57$^\dagger$                                       & 84.81                                   & 21.56$^\dagger$                                  \\
\multicolumn{1}{c|}{}                              & 512  & 14.16                 & 8.33                  & 6.58                  & 26$^\ddagger$~~                                         & 75.90$^\dagger$                                       & 168.96                                  & 47.64$^\dagger$                                  \\
\multicolumn{1}{c|}{}                              & 1024 & 38.19                 & 21.62                 & 16.89                 & 34.3$^\dagger$                                        & 83.12$^\dagger$                                       & 349.41                                  & 101.84$^\dagger$                                 \\
\multicolumn{1}{c|}{}                              & 2048 & 95.84                 & 53.03                 & 41.18                 & -~                                            & 363.90~                                       & 736.92                                  & -~                                       \\
\multicolumn{1}{c|}{}                              & 4096 & 230.45                & 124.95                & 96.62                 & -~                                            & 392.69~                                       & 1503.31                                 & -~                                       \\ \hline
\multicolumn{1}{c|}{\multirow{5}{*}{Energy (nJ)}}  & $N$: 256  & 0.80                  & 0.49                 & -                 & 0.144$^\ddagger$*                                      & 68.67$^\dagger$                                       & 570.60                                  & 2.15$^\dagger$                                   \\
\multicolumn{1}{c|}{}                              & 512  & 4.77                  & 2.67                 & -                 & 0.324$^\ddagger$*                                      & 75.90$^\dagger$                                       & 1179.52                                 & 5.28$^\dagger$                                   \\
\multicolumn{1}{c|}{}                              & 1024 & 13.86                 & 7.16                 & -                 & 0.868$^\dagger$*                                       & 83.12$^\dagger$                                       & 2483.77                                 & 12.52$^\dagger$                                  \\
\multicolumn{1}{c|}{}                              & 2048 & 36.68                 & 18.98                & -                & -~~~                                            & 363.60~                                       & 5273.07                                 & -~                                       \\
\multicolumn{1}{c|}{}                              & 4096 & 93.08                 & 48.93                & -                & -~~~                                            & 421.78~                                       & 10864.64                                & -~                                       \\ \hline
\multicolumn{9}{l}{\emph{Note.}~ 1. $\dagger$ indicates 16-bit bitwidth and $\ddagger$ 14-bit.} \\
\multicolumn{9}{l}{~~~~~~~  2. *very small memory, supporting only $N\leq$ 1024.}
\end{tabular}
    \vspace{-3mm}
\end{table*}

\subsection{Performance and Effect of Using Multiple Buffers}
\label{s-effectiveness}

\figref{buffer_sensitivity} compares performance of our PIM architecture under various values of $N$ (polynomial length) and $N_b$.  First  we note that without auxiliary buffers, there is no performance advantage even compared with a software execution, whereas even just one auxiliary buffer can improve performance by an order of magnitude.  

Moreover, adding more buffers gives very significant speed up of about $1.5 \sim 2.5\times$ depending on $N$.  
As expected, having multiple auxiliary buffers proves more effective when $N$ is larger, which is because at larger $N$, a bigger portion of runtime is accounted for by inter-row mapping and inter-row mapping benefits more from pipelining. 


\subsection{Sensitivity to Clock Frequency}
To see the effect of lower clock frequency we have varied the frequency from 300 MHz to 1200 MHz.
The computation time of the CU increases in proportion to the inverse of clock frequency, but the absolute latency of DRAM memory access time (in ns) is kept constant. 
The result is summarized in \figref{clock_sensitivity}.
Since most of the latency of NTT-PIM is due to DRAM memory operations, the performance of NTT-PIM is quite tolerable under lower frequencies, still achieving $3\sim 7\times$ speedup compared to CPU. The performance of NTT with long polynomial lengths tend to be more robust on lower frequencies, slowing down 1.65$\times$ only when the clock frequency drops by 4$\times$.

\subsection{Comparison with Previous Work}

\tabref{t-cmp} compares our NTT-PIM with previous PIM-based NTT accelerators as well as x86 CPU and FPGA, in terms of latency and power. 
Our NTT-PIM achieves speedup of minimum 1.7$\times$ up to 17$\times$ depending on the polynomial size. It is important to note that ours is much more flexible than some of the compared works.  For instance, CryptoPIM \cite{nejatollahi2020cryptopim} has a limitation that the modulo is fixed (a severe drawback for FHE, which runs multiple NTTs using different modulo values) and both CryptoPIM and MeNTT \cite{li2022mentt} limit the maximum polynomial size.  Ours has no such restriction.  Moreover, the power consumption of MeNTT \cite{li2022mentt} is very low, which is largely due to the fact that its maximum polynomial size is very small (1K).  While our NTT-PIM's latency increases exponentially as the polynomial length increases, it is expected of any scheme supporting arbitrary polynomial length.  After all, the number of operations increases as $O(N \log N)$. But it has also to do with the fact that longer polynomials require frequent row activations due to larger portion of inter-row mapping. 




\begin{figure}
  \myincludegraphics[width=\linewidth]{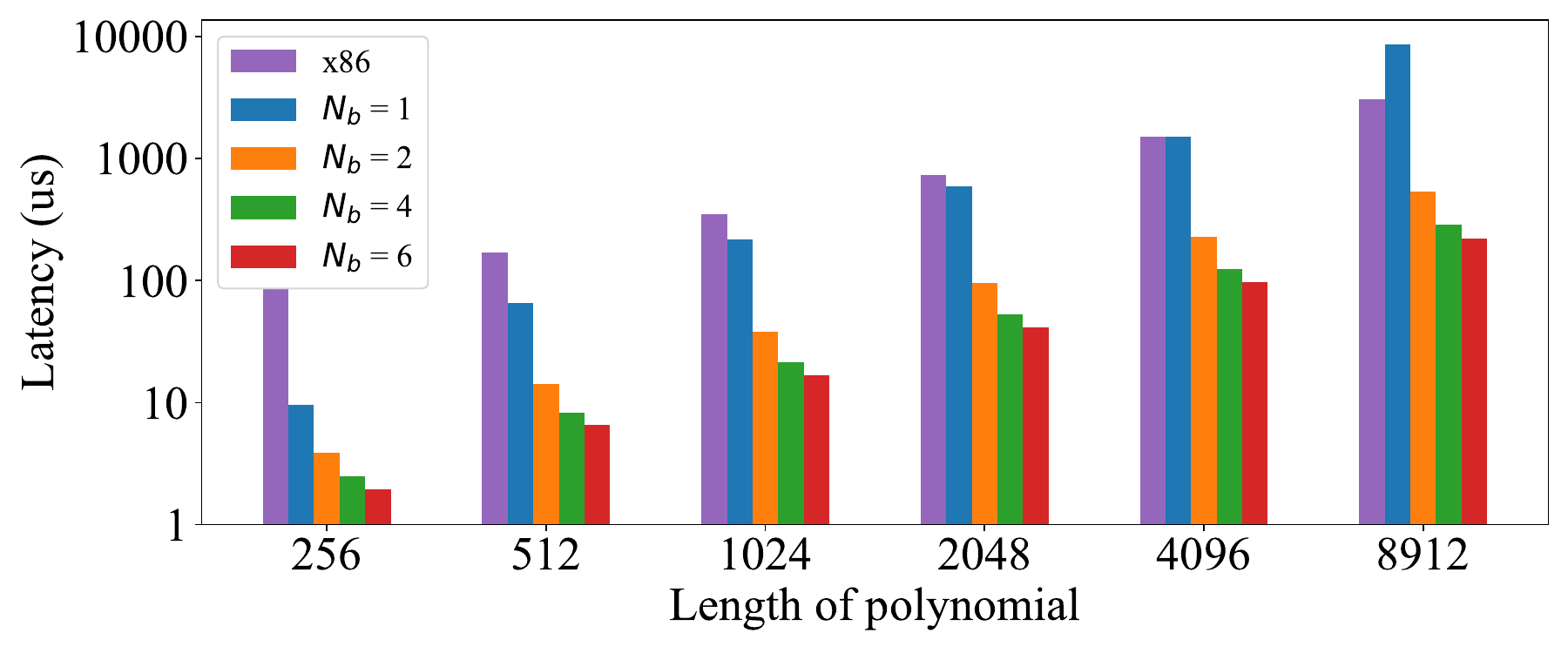}{Sensitivity to $N_b$, the number of buffers.} 
\end{figure}

\begin{figure}
  \myincludegraphics[width=\linewidth]{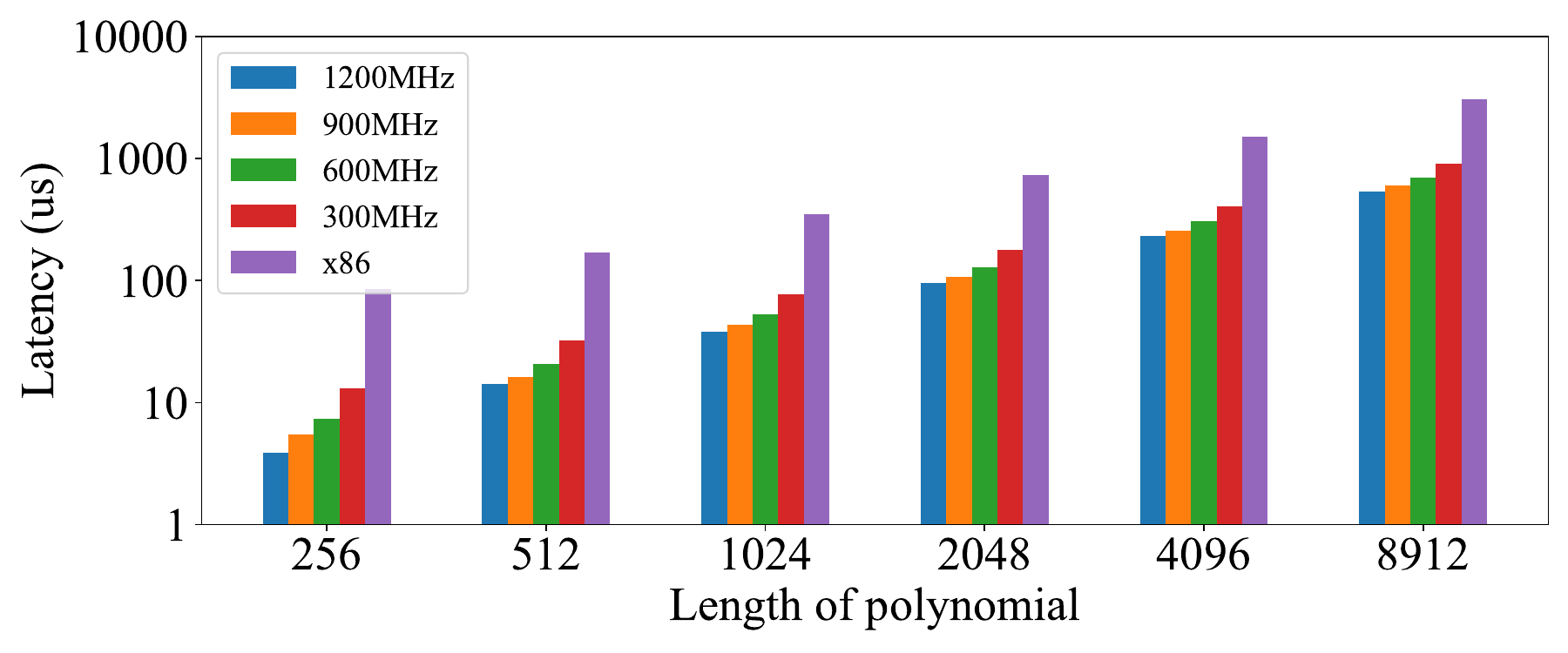}{Sensitivity to clock frequency ($N_b$ = 2).}
\end{figure}

\ignore{
Discussion.

Bank-level parallelism. 
MeNTT, CryptoPIM : modulo fixed, ours is flexible in terms of modulo. CKKS requires different modulo.  they all commonly use Barret ModMult, but they assume fixed modulo to simplify design. 
MeNTT: N too low. (max N is fixed? only supporting lower than certain N value; whereas ours has no such restriction on polynomial size) -- this is why their latency does not increase proportionally to N. 

Heax (FPGA): fixed modulo.  tens of BUs. 
F1 (FPGA).  hardware cost is very high. (at least compared to our solution).

Ours: selling point.
- hardware overhead extremely small
- very competitive performance and energy efficiency.
- power overhead: less than 1\% of DRAM access power
  area: same.

why our runtime grow faster than others.. 
1. N log N. -> should be more than 2x. 
2. relative portion of inter-row stages increases.  inter-row stages have more row activations than other regimes -> more increased latency.

inter-row: 

// remove --- MeNTT, CryptoPIM: polynomial mult. ok, but FHE not ok. why? key switch. 
}
\section{Conclusion}
We presented NTT-PIM, a novel PIM architecture supporting more flexible data movement within and across bank rows, and a mapping method for NTT functions.  To address the challenges of highly irregular memory access patterns and very restricted area budget, our solution exploits the characteristics of the algorithm, such as BU op-level scheduling, in-place update, and pipelining using multiple buffers.
Our experimental results demonstrates that even \textit{without modifying cell arrays}, PIM can provide state-of-the-art performance at very little area and power overhead for important functions such as NTT.

Our architecture and mapping is designed to support bank-level parallelism, and while we expect near-linear speed up as the number of banks increases, a more thorough investigation at the system level is left for future work.


\bibliographystyle{IEEEtran}
\bibliography{refs}

\begin{thebibliography}{10}
\providecommand{\url}[1]{#1}
\csname url@samestyle\endcsname
\providecommand{\newblock}{\relax}
\providecommand{\bibinfo}[2]{#2}
\providecommand{\BIBentrySTDinterwordspacing}{\spaceskip=0pt\relax}
\providecommand{\BIBentryALTinterwordstretchfactor}{4}
\providecommand{\BIBentryALTinterwordspacing}{\spaceskip=\fontdimen2\font plus
\BIBentryALTinterwordstretchfactor\fontdimen3\font minus
  \fontdimen4\font\relax}
\providecommand{\BIBforeignlanguage}[2]{{%
\expandafter\ifx\csname l@#1\endcsname\relax
\typeout{** WARNING: IEEEtran.bst: No hyphenation pattern has been}%
\typeout{** loaded for the language `#1'. Using the pattern for}%
\typeout{** the default language instead.}%
\else
\language=\csname l@#1\endcsname
\fi
#2}}
\providecommand{\BIBdecl}{\relax}
\BIBdecl

\bibitem{eet}
\BIBentryALTinterwordspacing
G.~Hilson, ``{AI} drives renewed interest in {PIM},'' \emph{EE Times}, 2021.
  [Online]. Available:
  \url{https://www.eetimes.com/ai-drives-renewed-interest-in-pim}
\BIBentrySTDinterwordspacing

\bibitem{fim}
N.~S. Kim, ``A journey to a commercial-grade processing-in-memory {(PIM)} chip
  development,'' in \emph{HPCA-27}, 2021, keynote speech.

\bibitem{9489313}
L.~Ke \emph{et~al.}, ``Near-memory processing in action: Accelerating
  personalized recommendation with {AxDIMM},'' \emph{IEEE Micro}, vol.~42,
  no.~1, pp. 116--127, 2022.

\bibitem{newton-isscc22}
S.~Lee \emph{et~al.}, ``A 1ynm {1.25V 8Gb, 16Gb}/s/pin {GDDR6}-based
  accelerator-in-memory supporting 1{TFLOPS} {MAC} operation and various
  activation functions for deep-learning applications,'' in \emph{ISSCC},
  vol.~65, 2022, pp. 1--3.

\bibitem{kwon2021fimdramm}
Y.-C. Kwon \emph{et~al.}, ``25.4 a 20nm 6gb function-in-memory dram, based on
  hbm2 with a 1.2tflops programmable computing unit using bank-level
  parallelism, for machine learning applications,'' in \emph{ISSCC}, vol.~64,
  2021, pp. 350--352.

\bibitem{mert2022anextensive}
A.~C. Mert \emph{et~al.}, ``An extensive study of flexible design methods for
  the number theoretic transform,'' \emph{IEEE Transactions on Computers},
  vol.~71, no.~11, pp. 2829--2843, 2022.

\bibitem{he2020newton}
M.~He \emph{et~al.}, ``{Newton}: A dram-maker’s accelerator-in-memory ({AiM})
  architecture for machine learning,'' in \emph{MICRO-53}, 2020, pp. 372--385.

\bibitem{10.1007/978-3-319-70694-8_15}
J.~H. Cheon \emph{et~al.}, ``Homomorphic encryption for arithmetic of
  approximate numbers,'' in \emph{Advances in Cryptology -- ASIACRYPT 2017},
  T.~Takagi \emph{et~al.}, Eds.\hskip 1em plus 0.5em minus 0.4em\relax Cham:
  Springer International Publishing, 2017, pp. 409--437.

\bibitem{riazi2020heax}
M.~S. Riazi \emph{et~al.}, ``{HEAX}: An architecture for computing on encrypted
  data,'' in \emph{ASPLOS}, 2020, p. 1295–1309.

\bibitem{samardzic2021f1}
N.~Samardzic \emph{et~al.}, ``F1: A fast and programmable accelerator for fully
  homomorphic encryption,'' in \emph{MICRO-54}, 2021, p. 238–252.

\bibitem{li2022mentt}
D.~Li \emph{et~al.}, ``{MeNTT}: A compact and efficient processing-in-memory
  number theoretic transform ({NTT}) accelerator,'' \emph{IEEE Transactions on
  Very Large Scale Integration (VLSI) Systems}, vol.~30, no.~5, pp. 579--588,
  2022.

\bibitem{nejatollahi2020cryptopim}
H.~Nejatollahi \emph{et~al.}, ``{CryptoPIM}: In-memory acceleration for
  lattice-based cryptographic hardware,'' in \emph{2020 57th ACM/IEEE Design
  Automation Conference (DAC)}, 2020, pp. 1--6.

\bibitem{chatterjee2017hpca}
N.~Chatterjee \emph{et~al.}, ``Architecting an energy-efficient dram system for
  gpus,'' in \emph{2017 IEEE International Symposium on High Performance
  Computer Architecture (HPCA)}, 2017, pp. 73--84.

\bibitem{cryptoeprint:2012/144}
J.~Fan \emph{et~al.}, ``Somewhat practical fully homomorphic encryption,''
  \emph{{IACR} Cryptol. ePrint Arch.}, p. 144, 2012.

\bibitem{10.1007/978-3-642-13190-5_1}
V.~Lyubashevsky \emph{et~al.}, ``On ideal lattices and learning with errors
  over rings,'' in \emph{Advances in Cryptology -- EUROCRYPT 2010}, H.~Gilbert,
  Ed.\hskip 1em plus 0.5em minus 0.4em\relax Berlin, Heidelberg: Springer
  Berlin Heidelberg, 2010, pp. 1--23.

\bibitem{Pppelmann2012TowardsEA}
T.~P{\"o}ppelmann \emph{et~al.}, ``Towards efficient arithmetic for
  lattice-based cryptography on reconfigurable hardware,'' in \emph{Progress in
  Cryptology -- LATINCRYPT 2012}, A.~Hevia \emph{et~al.}, Eds.\hskip 1em plus
  0.5em minus 0.4em\relax Berlin, Heidelberg: Springer Berlin Heidelberg, 2012,
  pp. 139--158.

\bibitem{10.1145/321450.321457}
M.~C. Pease, ``An adaptation of the fast {Fourier} transform for parallel
  processing,'' \emph{J. ACM}, vol.~15, no.~2, p. 252–264, apr 1968.

\bibitem{10.1016/S0167-8191(84)90413-7}
P.~N. Swarztrauber, ``{FFT} algorithms for vector computers,'' \emph{Parallel
  Comput.}, vol.~1, no.~1, p. 45–63, aug 1984.

\bibitem{Harris1977VectorRF}
D.~Harris \emph{et~al.}, ``Vector radix fast {Fourier} transform,'' in
  \emph{ICASSP '77. IEEE International Conference on Acoustics, Speech, and
  Signal Processing}, vol.~2, 1977, pp. 548--551.

\bibitem{pum16}
\BIBentryALTinterwordspacing
V.~Seshadri \emph{et~al.}, ``The processing using memory paradigm:in-dram bulk
  copy, initialization, bitwise {AND} and {OR},'' 2016. [Online]. Available:
  \url{https://arxiv.org/abs/1610.09603}
\BIBentrySTDinterwordspacing

\bibitem{6581570}
A.~Aysu \emph{et~al.}, ``Low-cost and area-efficient fpga implementations of
  lattice-based cryptography,'' in \emph{2013 IEEE International Symposium on
  Hardware-Oriented Security and Trust (HOST)}, 2013, pp. 81--86.

\bibitem{dramsim3}
S.~Li \emph{et~al.}, ``{DRAMsim3}: A cycle-accurate, thermal-capable dram
  simulator,'' \emph{IEEE Computer Architecture Letters}, vol.~19, no.~2, pp.
  106--109, 2020.

\bibitem{Montgomery1985ModularMW}
P.~L. Montgomery, ``Modular multiplication without trial division,''
  \emph{Mathematics of Computation}, vol.~44, pp. 519--521, 1985.

\bibitem{cacti-3dd}
K.~Chen \emph{et~al.}, ``{CACTI-3DD}: Architecture-level modeling for 3d
  die-stacked {DRAM} main memory,'' in \emph{2012 Design, Automation \& Test in
  Europe Conference \& Exhibition (DATE)}, 2012, pp. 33--38.

\end{thebibliography}

\end{document}